\documentstyle[12pt,a4]{article}

\newcommand{\be}{\begin{equation}}
\newcommand{\ee}{\end{equation}}
\newcommand{\unop}{1\!{\rm I}}
\def\comp{\hbox{{\rm C}\kern-.55em\hbox{{\rm I}}}}
\def\nat{\hbox{{\rm N}\kern-1em\hbox{{\rm I}}}}
\date{\ }
\title{Quasi exactly solvable matrix models in sl(n)}
\author{Y. Brihaye \\
Department of Mathematical Physics\\
University of Mons\\
Av. Maistriau, B-7000 MONS, Belgium.\\
       Piotr Kosinski \thanks{$^{\dagger}$ Work supported by grant $n^o$
KBN 2P03B07610}\\
Department of Theoretical Physics\\
University of Lodz\\
Pomorska 149/153, 90-236 Lodz, Poland}
\begin{document}
\begin{titlepage}
\maketitle
\thispagestyle{empty}
\begin{abstract}
We reconsider the quasi exactly solvable matrix models constructed recently
by R. Zhdanov. The 2$\times$2 matrix operators representing the algebra sl(2)
are generalized to matrices of arbitrary dimension and a similar construction
is achieved for the algebra sl(n).
\end{abstract}
\end{titlepage}
1. Recently Zhdanov [1],[2] considered the 
realizations of the sl(2) algebra in terms
of first order matrix differential operators. It reads
\be
Q_- = {d\over{dx}} \ \ , \ \ 
Q_0 = x{d\over dx} + A \ \ , \ \ 
Q_+ = x^2{d\over{dx}} + 2xA + B
\ee
\be
[Q_0,Q_{\pm}] = \pm Q_{\pm} \ \ \ , \ \ \
[Q_-,Q_+] = 2 Q_0
\ee
where $A$ and $B$ are (in general complex) $L\times L$ matices obeying
\be
[A,B] = B
\ee
This realization acts in the tensor product of $\comp\ ^L$ with the
linear space of smooth functions. 
In the following this vector space is denoted $V$.
Zhdanov posed and solved, in the case $L=2$, 
the problem of characterizing the invariant finite-dimensional 
subspaces of the representation space $V$.
\par In the present note we give the general characterization of such subspaces.
The main result is that they all basically arise from the Clebsch-Gordan
decomposition of the tensor product of the 
differential representation of spin $n/2$,
\be
q_- = {d\over{dx}}\quad , \quad 
q_0 = x{d\over{dx}}-{n\over 2}\quad, \quad 
q_+ = x^2{d\over{dx}} -nx \ \ ,
\ee
with the standard matrix representation of some spin $s$.
\par 2. Let us first consider the general form of two
 finite-dimensional operators $X,Y$ obeying
\be
\label{XY}
[X,Y] = Y
\ee
It follows from eq.~(\ref{XY}) that
\be
\label{XYMN}
(X-(\lambda+m))^nY^m = Y^m(X-\lambda)^n
\ee
for all integers $m,n\geq 0$.
\par Let us put $X$ in Jordan form. The space $R$ in which $X$ and $Y$ act
is spanned by the (generalized) eigenvectors of $X$. Let $\psi$ be some
(generalized) eigenvector of $X$ corresponding to the eigenvalue $\lambda$.
Then $Y\psi, Y^2\psi,...$ correspond to the eigenvalues $\lambda+1, \lambda+2,...$
and, if non vanishing, are linearly independent. This implies $Y^k\psi=0$
for some $k\leq N$, where $N$ is the dimension of the vector space under consideration.
Therefore $Y$ is nilpotent, $Y^K\equiv 0$ for some $K\leq N$. Note in passing
that if we apply this result to $X=-Q_0, Y=Q_-$, restricted to some 
finite-dimensional
subspace of $V$ we get at once the conclusion that such a subspace consists
of vectors with polynomial entries.
\par In order to get more information about the structure of
the operators $X$
and $Y$, let us define $\lambda$ to be an
 {\sl{isolated}} (generalized) eigenvalue
of $X$ if neither $\lambda-1$ nor $\lambda+1$ are (generalized) eigenvalues.
Let $R_0$ be the direct sum of all subspaces of $R$ which correspond to the
Jordan blocks of $X$ related to isolated eigenvalues. Then, due to 
eq.~(\ref{XYMN}),
$R_0$ is invariant under the action of $X$ and $Y$, $Y|_{R_0}=0$ and the whole
space $R$ is a sum of $R_0$ and of the invariant subspace $R_1$ corresponding
to the remaining Jordan blocks of $X$. Due to the definition of isolated
eigenvalues the set of remaining eigenvalues is a disjoint sum of finite
collection of chains $\Lambda_{\alpha}$ consisting of eigenvalues differing by one,
$\Lambda_{\alpha} = \lbrace \lambda_{\alpha}, \lambda_{\alpha}+1,\cdots,
\lambda_{\alpha}+n_{\alpha}-1 \rbrace,n_{\alpha} \in \nat$\ \ .
The subspace $R_1$ can be again written as $R_1=\oplus_{\alpha}R_{\alpha}$,
where $R_{\alpha}$ corresponds to the Jordan blocks related to $\lambda_{\alpha},
\lambda_{\alpha}+1,\cdots,\lambda_{\alpha}+n_{\alpha}-1$; each $R_{\alpha}$
is an invariant subspace under the action of $X$ and $Y$. 
\par
However, it is obviously
not the end of the story. The subspaces corresponding to some 
$\lambda_{\alpha}+L,
\lambda_{\alpha}+L+1$ can be (but not necessarily are) related by the action
of $Y$. Therefore each chain $\Lambda_{\alpha}$ can further be decomposed
into disjoint sum of subchains consisting of the eigenvalues having the
property that the subspaces corresponding to consecutive eigenvalues are related
by the action of $Y$ operator. Any subspace $R_{\alpha}$ can be correspondingly
decomposed in direct sum of invariant subspace corresponding to such subchains.
Therefore we have only to characterize such subspaces. Let $S$ be one of them.
One can write, according to the Jordan theorem,
\be
S= \oplus^K_{i=1} S_i
\ee
where $S_i$ is the generalized eigenspace corresponding to the eigenvalue
$\lambda_i = \lambda+(i-1)$ of $X$. The basis in $S_i$ is spanned by the
vectors $\psi_{i,j}$ , $j=1,\cdots,n_i$~, 
$n_i = {\rm dim} S_i$ such that 
\begin{eqnarray}
\psi_{i,j} &=&
(X-\lambda_i)^{j-1}\psi_{i,1} \ \ , \ \  j=1,\cdots, n_i \nonumber \\
(X-\lambda_i)^{n_i} \psi_{i,1} &=& 0
\end{eqnarray}
According to our definition of $S$ and to eq.~(\ref{XYMN}) we have
\begin{eqnarray}
Y S_i \subset S_{i+1}\quad , \quad YS_i &\not=& 0\quad , 
\quad i=1,\cdots,K-1 \nonumber\\
Y S_K &=& 0
\end{eqnarray}
\par
Let us describe in more detail the map 
$Y:S_i\rightarrow S_{i+1}$. First we note
that $Y\psi_{i,1}\not= 0$ for $i=1,\cdots, K-1$.
Indeed, if $Y\psi_{i,1}=0$,
then
\be
Y\psi_{i,j} = Y(X-\lambda_i)^{j-1}\psi_{i,1} = (X-(\lambda_i+1))^{j-1}Y
\psi_{i,1}=0
\ee
contrary to the definition of $S$. Therefore, by choosing an appropriate
normalization one can write
\be
Y\psi_{i,1} = \psi_{i+1,m_i}\quad , \quad 1\leq m_i \leq n_{i+1}
\ee
Then
\begin{eqnarray}
Y\psi_{i,j} &=& Y(X-\lambda_i)^{j-1}\psi_{i,1} = (X-(\lambda_i+1))^{j-1}
Y\psi_{i,1}\nonumber\\
&=& (X-\lambda_{i+1})^{j-1}\psi_{i+1,m_i} = 
\left\lbrace\begin{array}{cc}
\psi_{i+1,j+m_i-1} & {\rm if} \ \  j+m_i-1\leq n_{i+1}\\
0 &{\rm otherwise}
\end{array}\right.
\end{eqnarray}
The vectors $\psi_{i,n_i} \equiv (X-\lambda_i)^{n_i-1}\psi_{i,1}$ are the
eigenvectors of $X$. Therefore
\be
(X-(\lambda_i+1))Y\psi_{i,n_i} = Y(X-\lambda_i)\psi_{i,n_i} = 0
\ee
which implies either 
$Y\psi_{i,n_i} = 0$  or 
$Y\psi_{i,n_i} = \psi_{i+1,n_{i+1}}$.
So finally we conclude that $n_i+m_i-1\geq n_{i+1}$. 
Our subspace $S$ is
uniquely characterized by 
\begin{itemize}
\item an arbitrary complex number $\lambda$, 
\item an arbitrary natural number $K$, 
\item the sets of natural numbers
$n_i, \ i=1,\cdots, K, \ m_i, \ i=1,\cdots,K-1$ obeying : 
\begin{eqnarray}
1\leq m_i \leq n_{i+1} \ , \ i=1, \cdots \ , \  K-1 , \nonumber \\
 n_i+m_i-1\geq n_{i+1} \ ,\  i = 1,\cdots, K-1.
\end{eqnarray}
\end{itemize}
\par Let us note that our subspace $S$, being invariant under the action
of $X$ and $Y$, carries in general a not completely reducible representation of the
algebra (\ref{XY}). In what follows we will be interested in some special invariant
subspaces of $S$. Assume that $n_i+m_i - 1 = n_{i+1}, i=1, \cdots, K-1$ (if
necessary, we can decompose the set of eigenvalues $\lambda_i = \lambda + i-1$
into disjoint sum of subsets obeying these equalities and consider the
corresponding decomposition of $S$). Then we easily conclude that the subspace
of $S$ spanned by the eigenvectors $\phi_i \equiv \psi_{i,n_i}, i=1,\cdots,
K$ is invariant under the action of $X$ and $Y$. Moreover, $X$ and $Y$, when
restricted to this subspace, take the form :
\be
X = {\rm diag} (\lambda, \lambda+1, \ldots , \lambda+K-1)
\ee
\be
Y_{ij} =
\left\lbrace\begin{array}{cc}
1 & {\rm if} \ \ i=j+1  \\
0 &{\rm otherwise}
\end{array}\right.
\ee
\par
We shall call such invariant subspaces $sl(2)$ subspaces. They are the maximal
subspaces in which the operators $X,Y$ can be extended to the irreducible
representations of the $sl(2)$ algebra in the following sense. 
We take the traceless part of $X$
\begin{eqnarray}
X &=& {\rm diag}\ ( {1-K \over 2},{3-K \over 2}, \dots , {K-1 \over 2})
+ (\lambda + {K-1\over 2}) \unop \nonumber \\ 
  &\equiv& X_0 + (\lambda+ {K-1\over 2}) \unop
\end{eqnarray}
and we define the matrix $Z$ as
\be
Z_{ij} =
\left\lbrace\begin{array}{cc}
{i(i-K)\over 2} \ \  & {\rm if} \ \ j=i+1  \\
0 &{\rm otherwise}
\end{array}\right.
\ee
Then $X_0,Y,Z$ span the  sl(2) algebra :
\be
[X_0,Y] = Y\quad , \quad [X_0,Z]=-Z\quad , \quad [Z,Y] = 2X_0
\ee
It is further obvious that any eigenvector of $X$ belongs to a unique sl(2)
subspace; if it belongs to $R_0$, 
then the matrices $X_0$ and $Z$ are simply zero.
\par
3. Let us now come back to the algebra (2). 
We will use the results of our discussion
taking $X=A, Y=B$. Let $U \subset V$ be a finite-dimensional invariant subspace. The
sl(2) algebra is semi-simple so its finite-dimensional representations are
completely reducible. We can therefore take $U$ to be irreducible. Let $\psi$
be the lowest-weight vector, i.e. such that
\be
Q_-\psi = 0\quad ,\quad  Q_0 \psi = -{n\over 2}\psi
\ee
with $n$ integer. It follows then that $\psi$ is a constant eigenvector
of $A$ corresponding to the eigenvalue $-{n/2}$. Let $S$ be the sl(2)
subspace of $R$ containing $\psi$ and let $W$ be the tensor product of $S$
with the linear space of smooth functions of one variable. Obviously $W$
is invariant under the action of the algebra (1) and $W\cap U$ is nonempty
($\psi \in W\cap U$). Therefore $U = W\cap U \subset W$. Let us denote
the matrices $A,B$ restricted to $W$ also by $A,B$. According to the previous
discussion there exists a matrix $C$ such that
\be
A_0 = A-{1\over K} ({\rm Tr} A)\ \unop \equiv A-\mu \unop
\ee
together with $B$ and $C$ form the sl(2) algebra. We perform in $W$ the
following similarity transformation
\be
\tilde Q_- = e^{-xC} Q_- e^{xC} = {d\over {dx}} + C
\ee
\be
\tilde Q_0 = e^{-xC} Q_0 e^{xC} = (x{d\over {dx}} + \mu) + A_0
\ee
\be
\tilde Q_+ = e^{-xC} Q_+ e^{xC} = (x^2{d\over {dx}} + 2\mu x)+B
\ee
This transformation is a polynomial one ($C$ is nilpotent!). We see that
in $W$ there acts a product of two representations of sl(2) : the one spanned
by differential operators ${d\over{dx}}, x{d\over{dx}} + \mu, x^2{d\over{dx}}
+ 2\mu x$ and a matrix one. It is easy to see that it can contain a 
finite-dimensional
one only provided $\mu = {-m\over 2}$ for integer $m$. Therefore our
representation $U$ is obtained from the Clebsch-Gordan decomposition of the
product of finite-dimensional differential and matrix representations.
\par 
4. The structure described by eqs. (1) can be immediately generalized to
the $sl(N+1)$ algebra for any $N\geq 1$. Let us define the following generators
\begin{eqnarray}
\label{sln1}
J^k_0 &=& {\partial\over{\partial x_k}} \nonumber \\
J^0_k &=& -x_k (\sum^N_{j=1}x_j{\partial\over{x_j}} + C) -\sum^N_{j=1}x_j
A^j_k-B_k \nonumber \\
J^i_k &=& -x_k \partial_i - A_k^i -{1\over{N+1}} 
\delta^i_k (C-\sum^N_{j=1}A^j_j)
\end{eqnarray}
for $i,j,k = 1, \dots ,N$,
where $A^j_i, B_i$ and $C$ are the (in general complex) matrices obeying the
algebra
\begin{eqnarray}
\label{alg}
&[& A_i^j,A^l_k] = \delta^j_k A^l_i - \delta^l_i A^j_k \nonumber \\
&[& A_i^k,B_l] = \delta^k_l B_i \nonumber \\
&[& B_i,C] = B_i
\end{eqnarray}
the remaining commutators being vanishing. This algebra can be obtained by
taking a product of a differential and of a matrix representations of $sl(N+1)$, and then by
making a suitable similarity transformation. It is our guess that any
finite-dimensional subrepresentation of the algebra
(\ref{sln1}) can be obtained in this
way (as it was for the case $N=1$). In any case, due to the commutation rule
\be
[J^k_k,J^k_0] = J^k_0
\ee
we conclude again that $J^k_0$ when restricted to 
finite-dimensional subrepresentation
is nilpotent, i.e. such representation is necessarily a polynomial one.

\end{document}